\documentclass[11pt]{article}
\usepackage[margin=1in]{geometry}
\usepackage{graphicx}
\usepackage{epstopdf}
\usepackage{amssymb}
\usepackage{amsmath}
\usepackage{booktabs}
\usepackage{amsthm}
\usepackage[table]{xcolor}
\usepackage{stmaryrd}
\usepackage{mathabx}
\usepackage{mathrsfs}
\usepackage{latexsym}
\usepackage{rotating}
\usepackage{enumitem}
\usepackage[ruled,vlined,commentsnumbered,titlenotnumbered]{algorithm2e}
\usepackage{algorithmic}
\usepackage{url}
\usepackage{comment}
\usepackage{cite}
\usepackage{mathtools}
\usepackage{relsize}
\usepackage{exscale}

\excludecomment{comment}
\numberwithin{figure}{section}

\newtheorem{theorem}{Theorem}[section]
\newtheorem{lemma}[theorem]{Lemma}

\newtheorem{observation}[theorem]{Observation}

\theoremstyle{definition}

\newcommand{\ignore}[1]{}

\newcommand{\ceil}[1]{\lceil #1 \rceil}
\newcommand{\floor}[1]{\lfloor #1 \rfloor}

\newcommand{\dist}{\mathsf{dist}}
\newcommand{\polylog}{\operatorname{polylog}}

\newcommand{\DTW}{\textsf{DTW}}

\newcommand{\GED}{\textsf{GED}}

\newcommand{\cost}{\operatorname{cost}}
\newcommand{\ccost}{\operatorname{c-cost}}
\newcommand{\row}{\operatorname{r}}
\newcommand{\col}{\operatorname{c}}

\def\reals{{\mathbb R}}
\def \eps{{\varepsilon}}

\newcommand{\Frechet}{Fr\'echet}

\title{Dynamic Time Warping and Geometric Edit Distance:\\ Breaking the Quadratic Barrier\thanks{Work on this paper has been supported
by Grant 892/13 from the Israel Science Foundation, 
by Grant 2012/229 from the U.S.-Israeli Binational Science Foundation,
by the Israeli Centers of Research Excellence (I-CORE)
program (Center No.~4/11),
and by the Hermann Minkowski--MINERVA Center for Geometry at Tel Aviv
University.}}

\author{Omer Gold\thanks{Blavatnik School of Computer Science, Tel Aviv University, Tel Aviv 69978,
Israel; {\tt omergold@post.tau.ac.il}} 
\and Micha Sharir\thanks{Blavatnik School of Computer Science, Tel Aviv University, Tel Aviv 69978,
Israel; {\tt michas@post.tau.ac.il}}} 

\begin{document}

\sloppy


\maketitle

\begin{abstract}
Dynamic Time Warping (DTW) and Geometric Edit Distance (GED) are basic similarity measures between curves
or general temporal sequences (e.g., time series) that are represented as sequences of points in some metric space $(X, \dist)$.
The DTW and GED measures are {\em massively} used in various fields of computer science, computational biology, and engineering. 
Consequently, the tasks of computing these measures are among the core problems in P.
Despite extensive efforts to find more efficient algorithms, 
the best-known algorithms for computing the DTW or GED between two sequences of points in $X = \reals^d$ are long-standing dynamic programming algorithms that require quadratic runtime,
even for the one-dimensional case $d = 1$, which is perhaps one of the most used in practice.

In this paper,
we break the nearly 50 years old quadratic time bound for computing DTW or GED between two sequences of $n$ points in $\reals$,
by presenting deterministic algorithms that run in $O\left( n^2 / \log\log n \right)$ time.
Our algorithms can be extended to work also for higher dimensional spaces $\reals^d$, for any constant $d$,
when the underlying distance-metric $\dist$ is polyhedral (e.g., $L_1, L_\infty$).

\end{abstract}


\section{Introduction}\label{sec:DTW}
Dynamic Time Warping (DTW) and Geometric Edit Distance (GED) are basic similarity measures between curves
or general temporal sequences (e.g., time series) that are represented as sequences of points in some metric space $(X, \dist)$.
The DTW and GED measures are massively used in various fields of computer science, computational biology, and engineering.
Consequently, the tasks of computing these measures are among the core problems in P.
Despite extensive efforts to find more efficient algorithms, 
the best-known algorithms for computing the DTW or GED between two sequences of points in $X = \reals^d$ are long-standing dynamic programming algorithms that require quadratic runtime,
even for the one-dimensional case $d = 1$, which is perhaps one of the most used in practice.

In this paper, we present deterministic algorithms that run in $O\left( n^2 / \log\log n \right)$ time,
for computing DTW or GED between two sequences of $n$ points in $\reals$.
This result breaks the nearly 50 years old quadratic time bound for this problems.
Our algorithms can be extended to work also for higher dimensional spaces $\reals^d$, for any constant $d$,
when the underlying distance-metric $\dist$ is polyhedral (e.g., $L_1, L_\infty$).

\subsection{Problem Statements}\label{DTW:ProblemStatement}
Let $A= (p_1,\ldots, p_n)$ and $B= (q_1, \ldots, q_{m})$ be two sequences of points (also referred to as curves) in some metric space $(X, \dist)$.
A {\em coupling} $C=(c_1,\ldots,c_k)$ between $A$ and $B$ is an ordered sequence of distinct pairs of points from $A \times B$,
such that $c_1=(p_1, q_1)$, $c_k=(p_n, q_m)$, and
\[ c_r=(p_i, q_j)\Rightarrow c_{r+1}\in \bigl\{(p_{i+1}, q_j),\, (p_i, q_{j+1}),\, (p_{i+1}, q_{j+1})  \bigr\}, \]
for $r < k$.
The \DTW{}-distance between $A$ and $B$ is 
\begin{equation}\label{eq:dtw}
\mathsf{dtw}(A,B) = \min_{C:\,coupling} \biggl\{\sum_{(p_i, q_j) \in C}{\dist(p_i,\, q_j)}\biggr\}.
\end{equation}
A coupling $C$ for which the above sum is minimized is called an {\em optimal~coupling}.
The \DTW{} problem is to compute $\mathsf{dtw}(A,B)$, and sometimes also an optimal coupling $C$.

\begin{figure}[t]
	\centering
	\includegraphics[scale = 0.5]{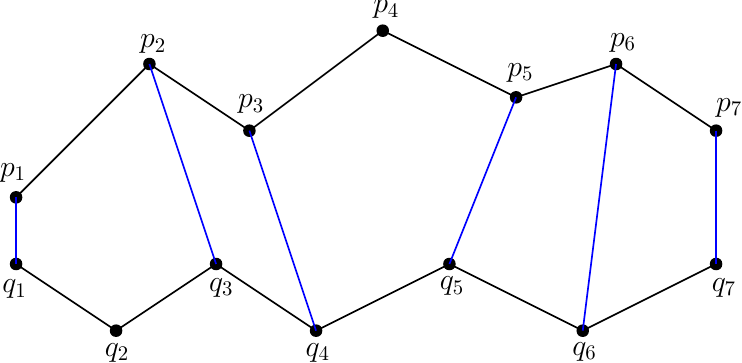}
	\caption{Example of a monotone matching (in blue) between two polygonal curves (represented by point-sequences) in the plane.}
	\label{fig:matching}
\end{figure}

A {\em monotone matching} $\mathcal{M} = \{ m_1,\ldots,m_k \}$ between $A$ and $B$ is a set of pairs of points from $A \times B$,
such that any two pairs $(p_i, q_j),\, (p_{i'}, q_{j'}) \in  \mathcal{M}$ satisfy that  $i < i'$ iff $j < j'$. 
This also implies that each point in $A$ is matched with at most one point in $B$ and vice versa (possibly some points in $A\cup B$ do not appear in any pair of the matching); see Figure~\ref{fig:matching} for an illustration.
Note the difference from coupling (defined above), which covers all points of $A \cup B$ and a point can appear in multiple pairs of the coupling.
The cost of $\mathcal{M}$ is defined to be the sum of all the distances between the points of each pair in $\mathcal{M}$,
plus a {\em gap} penalty parameter $\rho \in \reals$, for each point in $A \cup B$ that does not appear in any pair of $\mathcal{M}$.

The Geometric Edit Distance (\GED{}) between $A$ and $B$ is
\begin{equation}\label{eq:ed}
\mathsf{ged}(A,B) = \min_{\mathcal{M}} \biggl\{ \Big(\sum_{(p_i, q_j) \in \mathcal{M}}{\dist(p_i,\, q_j) \Big)} + \rho\left(n + m -2|\mathcal{M}|\right) \biggr\},
\end{equation}
where the minimum is taken over all sets of monotone matchings $\mathcal{M}$ in the complete bipartite graph $A \times B$.
A monotone matching $\mathcal{M}$ for which the above sum is minimized is called an {\em optimal~matching}.
The \GED{} problem is to compute $\mathsf{ged}(A,B)$, and sometimes also an optimal matching.
More sophisticated gap
penalty functions have been proposed~\cite{DEKM98}, but for this presentation, we focus on the standard linear gap penalty
function, although our presented algorithm supports more complex gap penalty,
such as taking $\rho$ to be a linear function in the coordinates of the points of $A\cup B$.
By tuning $\rho$ correctly, meaningful matchings can be computed even when
faced with outlier points that arise from measurement errors or short deviations in otherwise
similar trajectories.

The \DTW-distance and \GED{} are massively used in dozens of applications, such as speech recognition, geometric shape matching,
DNA and protein sequences, protein backbones, matching of time series data, GPS, video and touch screen authentication trajectories, music signals, and countless data mining applications;
see~\cite{731723, DeLuca2012, Efrat2007, 10.2307/2242521, Keogh2000, Keogh2005, Muller2007, RE05, Keogh1999} for some examples.

The best-known worst-case running times for solving \DTW{} or \GED{} are given by long-standing simple dynamic programming algorithms that require $\Theta(nm)$ time.
We review the standard quadratic-time \DTW{} and \GED{} algorithms in Section~\ref{sec:quadratic} and~\ref{sec:ed}, respectively. 

\DTW{} was perhaps first introduced as a speech discrimination method~\cite{Vintsyuk1968} back in the 1960's.
\GED{} is a natural extension of the well-known string version of Edit Distance, however,
the subquadratic-time algorithms for the string version do not seem to extend to \GED{} (see below).

A popular setting in both theory and practice is the one-dimensional case $X=\reals$ (under the standard Euclidean distance $\dist(x,y) = |x-y|$). 
Even for this special case, no subquadratic-time algorithms have been known.
We consider this case throughout most of the paper.

\section{Summary of Our Results and Related Works}

\paragraph*{Prior Results.}
Since no subquadratic-time algorithm is known for computing \DTW{},
a number of heuristics were designed to speed up its exact computation in practice; see Wang {\em et al.}~\cite{Wang2013} for a survey.
Very recently, Agarwal {\em et al.}~\cite{AFPY16} gave a near-linear approximation scheme for computing \DTW{} or \GED{} for a restricted, although quite large, family of curves.

Recently, Bringmann and K{\"{u}}nnemann~\cite{BK15} proved that \DTW{} on one-dimensional point sequences whose elements are taken from $\{0,\, 1,\, 2,\, 4,\, 8\} \subset \reals$
has no $O(n^{2-\Omega(1)})$-time algorithm, unless SETH fails. They proved a similar hardness result also for Edit Distance between two binary strings,
improving the conditional lower bound of Backurs and Indyk~\cite{BI15}.
This line of work was extended in a very recent work by Abboud {\em et al.}~\cite{Abboud16}, and Abboud and Bringmann~\cite{AB18},
where they show that even a sufficiently large $\polylog(n)$-factor
improvement over the quadratic-time upper bound of similar quadratic matching problems, may lead to 
major consequences, such as faster Formula-SAT algorithms, and new circuit complexity lower bounds. 

Masek and Paterson~\cite{MMP80} showed that Edit Distance between two strings of length at most $n$ over an $O(1)$-size alphabet can be solved in $O(n^2 / \log n)$ time.
More recent works~\cite{BILLE2008486, Grabowski2016} managed to lift the demand for $O(1)$-size alphabet and retain a subquadratic-time bound by making a better use of the word-RAM model. 
However, these works do not seem to extend to \GED{}, especially not when taking sequences of points with arbitrary real coordinates.
In the string version, the cost of replacing a character is fixed (usually $1$), hence, we only need to detect that two characters are not identical in order to compute the replacement cost,
unlike in \GED{}, where the analogous cost for two matched points is taken to be their distance, under some metric.

\paragraph*{Our Results and Related Works.}
Efforts for breaking the quadratic time barrier for basic similarity measures between curves and point-sequences were recently stimulated by the result of Agarwal {\em et al.}~\cite{AAKS14}
who showed that the discrete \Frechet{} distance can be computed in $O(n^2 / \log n)$ time.
Their algorithm for (discrete) \Frechet{} distance does not extend to \DTW{} or \GED{},
as the formula for the (discrete) \Frechet{} distance uses the max function over distances between pairs of points,
while the formulas for \DTW{} and \GED{} involve their sum.
As a result, the \Frechet{} distance is effectively determined by a single pair of sequence elements, which fits well into the use of the Four-Russians technique~\cite{4Russians},
while the \DTW{} and \GED{} are determined by many pairs of elements.
This makes our algorithms much more subtle,
involving a combination and extension of techniques from computational geometry and graph shortest paths.

To simplify the presentation,
we present our results only for the ``balanced'' case $m=n$;
extending them to the general case $m \leq n$ is easy.
The standard $\Theta(mn)$-time algorithm is superior to our solution only when $m$ is subpolynomial in $n$.

Our results are stated in the following theorems.
\begin{theorem}\label{thm:first}
	Given two sequences $A= (p_1,\ldots, p_n)$ and $B= (q_1, \ldots, q_{n})$, each of $n$ points in $\reals$,
	the \textnormal{\DTW{}}-distance $\mathsf{dtw}(A,B)$ (and optimal coupling), or the \textnormal{\GED{}} $\mathsf{ged}(A,B)$ (and optimal matching) can be computed by a deterministic algorithm in $O(n^2 / \log\log n)$ time.
\end{theorem}
Theorem~\ref{thm:first} gives the very first subquadratic-time algorithm for solving \DTW{},
breaking the nearly 50 years old $\Theta(n^2)$ time bound~\cite{Vintsyuk1968}.
We present the improved algorithm for \DTW{} in Section~\ref{sec:dtw}.
In Section~\ref{sec:extended} we extend our algorithm to give a more general result,
which supports high-dimensional polyhedral metric spaces, as stated in Theorem~\ref{thm:second} given below.
In Section~\ref{sec:ed} we extend our algorithm to obtain a subquadratic solution for \GED{}.
\begin{theorem}\label{thm:second}
	Let $A= (p_1,\ldots, p_n)$ and $B= (q_1, \ldots, q_{n})$ be two sequences of $n$ points in $\reals^d$, where $d$ is a constant and the underlying distance-metric is polyhedral\footnote{That is, the underlying metric is induced by a norm, whose unit ball is a symmetric convex polytope with $O(1)$ facets (e.g., $L_1$, $L_\infty$).}.
	Then $\mathsf{dtw}(A,B)$ (and optimal coupling), or $\mathsf{ged}(A,B)$ (and optimal matching) can be computed by a deterministic algorithm in
	$O(n^2 / \log\log n)$ time.
\end{theorem}

\section[Preliminaries, Tools, and the Quadratic Time DTW Algorithm]{Preliminaries, Tools, and the Quadratic Time \DTW{} Algorithm}
Throughout this paper, we view matrices with rows indexed in increasing order from bottom to top and columns indexed in increasing order from left to right,
so for example, $M[0,0]$ corresponds to the value of the leftmost-bottom cell of a matrix $M$.

In Fredman's classic 1976 articles on the decision tree complexity of $(\min, +)$-matrix multiplication~\cite{F76}, and on sorting $X + Y$~\cite{Fredman76},
he often uses the simple observation that $a + b < a' + b'$ iff $a - a' < b'-b$. This observation is usually referred to as {\em Fredman's trick}.
In our algorithm, we will often use the following extension of Fredman's trick.
\begin{align}\label{eq:Fredman}
\begin{split}
a_1 - b_1 +\cdots + a_r - b_r &< a'_1 - b'_1 +\cdots + a'_t - b'_t \\
{\textrm {if and}} &\text{ }{\textrm {only if}} \\
a_1 + \cdots + a_r -a'_1 - \cdots -a'_t &< b_1 + \cdots + b_r - b'_1 - \cdots -b'_t.
\end{split}
\end{align}

Our algorithm uses a geometric domination technique, based on the following algorithm of Chan~\cite{Chan08}. 
\begin{lemma}[Chan~\cite{Chan08}]\label{lem:redblue}
	Given a finite set $P= \{p_1,\ldots, p_n\}$ of points in $\mathbb{R}^d$ such that each point is colored red or blue, one can report
	all pairs $(i,j)\in [n]^2$, such that $p_i$ is red, $p_j$ is blue, and $p_i[k] > p_j[k]$ for every $k\in [d]$,
	in time  $O(c_\eps^d |P|^{1+\eps} + K)$, where $K$ is the output size,
	$\eps \in (0,1)$ is an arbitrary prespecified parameter, and $c_\eps = 2^\eps/(2^{\eps}-1)$.
\end{lemma}
Throughout the paper, we invoke Lemma~\ref{lem:redblue} many times, with $\eps=1/2, c_\eps \approx 3.42,$ and
$d = \delta \log n$, where $\delta >0$ is a sufficiently small constant, chosen
to make the overall running time of all the invocations dominated by the total output size; see below for details.

We denote by $[N] = \{1,\ldots, \ceil{N} \}$, the set of the first $\ceil{N}$ natural numbers, for any $N\in \reals^+$.

Throughout this paper, we sometimes refer to a square matrix as a {\em box}.

Our model of computation is a simplified 
Real RAM model, in which ``truly real'' numbers are subject to only two unit-time operations: addition and comparison.
In all other respects, the machine behaves like a $w=O(\log n)$-bit word RAM with the standard repertoire of unit-time 
$AC^0$ operations, such as bitwise Boolean operations, and left and right shifts.

\subsubsection*{The Quadratic Time \DTW{} Algorithm}\label{sec:quadratic}
We give an overview of the standard dynamic programming algorithm for computing the \DTW{}-distance between two sequences of $n$ points in $\reals$, which requires quadratic time~\cite{Vintsyuk1968}.
This algorithm can be easily extended to return also the optimal coupling (see below).
In Section~\ref{sec:ed} we overview a ``similar in principle'' algorithm for solving \GED{}.
\smallskip

We are given as input two sequences $A= (p_1,\ldots, p_n)$ and $B= (q_1, \ldots, q_{n})$ of $n$ points in $\reals$.
(The algorithm below can be (trivially) modified to handle sequences of different lengths.)
\setdescription{leftmargin=1.5cm,labelindent=\parindent}
\begin{description}\label{alg:DTWquadratic}
	\addtolength{\itemsep}{-0.5\baselineskip}
	\item[1.$\;$] Initialize an $(n+1)\times (n+1)$ matrix $M$ and set $M[0,0]:=0$.
	\item[2.$\;$] For each $\ell \in [n]$
	\item [2.1.$\;\;$] $M[\ell, 0]:= \infty$, $M[0, \ell]:= \infty$. 
	\item[3.$\;$] For each $\ell\in [n]$, 
	\item[3.1.$\;\;$] For each $m\in [n]$,
	\item[3.1.1$\;\;\;$] 	$M[\ell,m] := \bigl|p_\ell - q_m\bigr| + \min \Bigl\{M[\ell-1,m],\, M[\ell,m-1],\, M[\ell-1,m-1] \Bigr\}$.
	\item[4.$\;$] Return $M[n,n]$.
\end{description}
The optimal coupling itself can also be retrieved, at no extra asymptotic cost,
by the standard technique of maintaining pointers from each $(\ell, m)$ to the preceding position
\[
(\ell', m')\in \left\{ (\ell-1, m),\, (\ell, m-1),\, (\ell-1, m-1) \right\}
\]
through which $M[\ell , m]$ is minimized.
Tracing these pointers backwards from $(n, n)$ to $(0, 0)$ and reversing these links yields the desired optimal coupling.

\section{Dynamic Time Warping in Subquadratic Time}\label{sec:dtw}
As above, the input consists of two sequences $A= (p_1,\ldots, p_n)$ and $B= (q_1, \ldots, q_{n})$ of $n$ points in~$\reals$. 
Our algorithm can easily be modified to handle the case where $A$ and $B$ have different lengths.

\subsection*{Preparations}
We fix some (small) parameter $g$, whose value will be specified later; for simplicity, we assume that $\frac{n}{g-1}$ is an integer.
We decompose $A$ and $B$ into $s = \frac{n}{g-1}$ subsequences $A_1,\ldots,A_{s}$, and $B_1,\ldots, B_{s}$, such that  
for each $i,j\in \{2,\ldots,s\}$, each of $A_i$ and $B_j$ consists of $g-1$ consecutive elements of the corresponding sequence,
prefixed by the last element of the preceding subsequence.
We have that $A_1$ and $B_1$ are both of size $g-1$, each $A_i$ and $B_j$ is of size $g$, for $i,j\in \{2,\ldots,s\}$,
and each consecutive pair $A_i,\, A_{i+1}$ or $B_j,\, B_{j+1}$ have one common element.

For each $i, j \in [s]$, denote by $D_{i,j}$ the {\em all-pairs-distances matrix} between points from $A_i$ and points from $B_j$;
specifically, $D_{i,j}$ is a $g \times g$ matrix (aka a {\em box}, see below for the cases $i=1$ or $j=1$) such that for every $\ell, m \in [g]$, 
\[ 
D_{i,j}[\ell,m] = \bigl |A_i(\ell) - B_j(m) \bigr|.
\] 
For all $i\in[s]$, we add a leftmost column with $\infty$ values to each box $D_{i, 1}$,
and similarly, we add a bottommost row with $\infty$ values to each box $D_{1,i}$.
In particular, $D_{1,1}$ is augmented by both a new leftmost column and a new bottommost row. The common element $D_{1,1}[0,0]$ of this row and column is set to $0$.
Overall, we have $s^2 = \left(\frac{n}{g-1}\right)^2$ boxes $D_{i,j}$, all of size $g \times g$.

We define a {\em staircase path} $P$ on a $g \times g$ matrix $D_{i,j}$ as a sequence of positions from $[g]\times [g]$ that form a monotone staircase structure,
starting from a cell on the left or bottom boundary and ending at the right or top boundary,
so that each subsequent position is immediately either to the right, above, or above-right of the previous one. Formally, by enumerating the path positions as 
$P(0), \ldots, P(t^*)$, we have
$P(t+1) \in \{ P(t) + (0,1), P(t) + (1,0), P(t) + (1,1)  \} $, for each $t \in \{0,\ldots, t^* -1\}$.
The path starts at some point $P(0)= (\cdot ,1)$ or $(1, \cdot)$,
and ends at some $t^*$ (not necessarily the first such index) for which $P(t^*) = (\cdot, g)$ or $(g, \cdot)$.
Note that $t^*$ can have any value in $[2g-2]$.
The number of possible monotone staircase paths in a box $D_{i,j}$ is trivially bounded by $O(g^2 3^{2g-2})$,
and the following more careful reasoning improves this bound to $O(3^{2g})$.
Each staircase path can be encoded by its first position, followed by its sequence of moves, where each move is in one of the directions up/right/up-right. Thus, the number of staircase paths that start at some position $(r,1)$ (resp. $(1,r)$) at the left (resp. bottom) boundary is bounded by $3^{2g-1-r}$. 
Thus, the total number of staircase paths that start at the left or the bottom boundary is bounded by
\[
2 \sum^{g}_{r=1}{3^{2g-1-r}} = O(3^{2g}).
\]

We define the {\em cost} of a staircase path $P$ in a box $D_{i,j}$ by 
\[
\cost_{i,j}(P) = \sum^{t^*}_{t=1} D_{i,j}(P(t)).
\]
(For technical reasons, that will become clear in the sequel, we generally do not include the first position $P(0)$ of the path in evaluating its cost,
except in the boxes $D_{i,1}$ and $D_{1,j}$ for all $i,j \in [s]$.)
To ease the presentation, in the algorithm that follows, we assume (or ensure) that no two distinct paths in a box $D_{i,j}$ have the same cost.
This will be the case if we assume that the input sequences are in sufficiently general position.
In Section~\ref{subsec:lifting} we will show how this assumption can be completely removed, by adding a few additional steps to the preprocessing stage of the algorithm, without increasing its asymptotic time bound.

We denote by $L$ the set of {\em positions} in the left and bottom boundaries of {\em any} box $D_{i,j}$,
and by $R$ the set of positions in the right and top boundaries (note that $L$ and $R$ have two common positions).
Given a starting position $v \in L$, and an ending position $w \in R$,
we denote by $S(v,w)$ the set of all staircase paths $P_{v,w}$ that start at $v$ and end at $w$ (if there is no staircase path between $v$ and $w$, then $S(v,w)= \emptyset$).
We say that $P^*_{v,w} \in S(v,w)$ is the {\em shortest path} between $v$ and $w$ in $D_{i,j}$ iff
\[
\cost_{i,j}\left(P^*_{v,w}\right) = \min_{P_{v,w} \in S(v,w)} \left\{\cost_{i,j}\left(P_{v,w}\right) \right\}.
\]
Note that according to our general position assumption, the shortest path between $v$ and $w$, within a given box, is {\em unique}.

\subsection*{First Stage: Preprocessing}
The first stage of our algorithm is to construct a data structure in subquadratic time (and storage),
such that for each box $D_{i,j}$, and for each pair of positions $(v,w)\in L \times R$,
we can retrieve the shortest path $P^*_{v,w}$ in $D_{i,j}$ and $\cost_{i,j}(P^*_{v,w})$ in $O(1)$ time, when such a path exists (i.e., when $S(v,w)$ is nonempty).

The algorithm enumerates all $(2g-1)^2$ pairs of positions $(v,w)$ in a $g \times g$ matrix (box) such that $v\in L$ and $w \in R$,
discarding pairs that cannot be connected by a monotone staircase path,
and referring to the surviving pairs as {\em admissible}.
Again, we simplify the notation by upper bounding this quantity by $4g^2$.
For each such admissible pair $(v,w) \in L\times R$, we also enumerate {\em every} possible staircase path in $S(v,w)$ as $P_{v,w}\,:\, [t^*]\rightarrow [g] \times [g]$,
where we write $P_{v,w} = \left(P^{\row}_{v,w}, P^{\col}_{v,w}\right)$ as a pair of
row and column functions $P^{\row}_{v,w},\, P^{\col}_{v,w} \,:\, [t^*]\rightarrow [g]$, so that $P_{v,w}(k) = \left(P^{\row}_{v,w}(k), P^{\col}_{v,w}(k) \right)$,
for each $k \in [t^*]$.
(Note that $t^*$ is a path-dependent parameter, determined by $v,\,w$ and the number of diagonal moves in the path.)
In total, there are $O(3^{2g})$ possible staircase paths $P_{v,w}$ (for all admissible pairs $(v,w)\in L\times R$ combined), which we enumerate.
The above enumerations are done using a natural lexicographic order, which induces a total order on the $< 4g^2$ admissible pairs of positions of $L \times R$, 
and for each such pair $(v,w)$, a total order on all possible staircase paths $P_{v,w}\in S(v,w)$.

Given two staircase paths $P_{v,w}$ and $P'_{v,w}$ with the same starting and ending positions $v,w$ in a box $D_{i,j}$,
we want to use the extended Fredman trick (as in~\eqref{eq:Fredman}) to compare $\cost_{i,j}\left(P_{v,w}\right)$ with $\cost_{i,j}\left(P'_{v,w}\right)$, 
by comparing two expressions such that one depends on points from $A_i$ only and the other depends on points from $B_j$ only.
Suppose that $P_{v,w} = \left( (\ell_1, m_1), \ldots, (\ell_r, m_r)\right)$ and $P'_{v,w} = \left( (\ell'_1, m'_1), \ldots, (\ell'_t, m'_t)\right)$
(note that $(\ell_r, m_r) = (\ell'_t, m'_t) = w$, since both paths end at $w$, and that we ignore the common starting positions $(\ell_0, m_0) = (\ell'_0, m'_0) = v$).
We have
\[\cost_{i,j}\left(P_{v,w}\right) = \bigl|A_i(\ell_1) - B_j(m_1)\bigr| + \cdots +\bigl|A_i(\ell_r) - B_j(m_r) \bigr|,\]
and
\[\cost_{i,j}\left(P'_{v,w}\right) = \bigl|A_i(\ell'_1) - B_j(m'_1)\bigr| + \cdots +\bigl|A_i(\ell'_t) - B_j(m'_t) \bigr|,\]
and we want to test whether, say, $\cost_{i,j}\left(P_{v,w}\right) < \cost_{i,j}\left(P'_{v,w}\right)$ (recall that we assume that equalities do not arise), 
that is, testing whether
\begin{equation}\label{eq:abs}
\bigl|A_i(\ell_1) - B_j(m_1)\bigr| + \cdots +\bigl|A_i(\ell_r) - B_j(m_r) \bigr| < \bigl|A_i(\ell'_1) - B_j(m'_1)\bigr| + \cdots +\bigl|A_i({\ell'_t}) - B_j(m'_t) \bigr|.
\end{equation}
The last term in each side of~(\ref{eq:abs}) is actually unnecessary, since they are equal.
In order to transform this inequality into a form suitable for applying the extended Fredman trick~\eqref{eq:Fredman},
we need to replace each absolute value $\left| x \right|$ by either $+x$ or $-x$, as appropriate.
To see what we are after, assume first that the expressions $A_i(\ell_k) - B_j(m_k)$ and $A_i(\ell'_k) - B_j(m'_k)$ are all positive, 
so that (\ref{eq:abs}) becomes  
\begin{equation*}
A_i(\ell_1) - B_j(m_1) + \cdots + A_i(\ell_r) - B_j(m_r) < A_i(\ell'_1) - B_j(m'_1) + \cdots + A_i({\ell'_t}) - B_j(m'_t).
\end{equation*}
By~(\ref{eq:Fredman}) we can rewrite this inequality as
\[
A_i(\ell_1) + \cdots + A_i(\ell_r) - A_i(\ell'_1) - \cdots - A_i(\ell'_t) < B_j(m_1) + \cdots + B_j(m_r) - B_j(m'_1) - \cdots - B_j(m'_t),
\]
which can be written as
\begin{align}
A_i(P^{\row}_{v,w}(1)) + \cdots + A_i(P^{\row}_{v,w}(r)) &- A_i(P'^{\row}_{v,w}(1)) - \cdots - A_i(P'^{\row}_{v,w}(t))  \label{alpha}\\
< B_j(P^{\col}_{v,w}(1)) + \cdots + B_j(P^{\col}_{v,w}(r)) &- B_j(P'^{\col}_{v,w}(1)) - \cdots - B_j(P'^{\col}_{v,w}(t))	\label{beta}.
\end{align}
If $P_{v,w} = P^*_{v,w}$ (i.e., if $P_{v,w}$ is the shortest path from $v$ to $w$) in $D_{i,j}$ then the inequality above holds for all pairs $(P_{v,w},\, P'_{v,w})$,
where $P'_{v,w} \in S(v,w)$ is any other staircase path between $v$ and $w$.

For each admissible pair of positions $(v,w)\in L \times R$, we choose some staircase path $P_{v,w}$ as a candidate for being the shortest path from $v$ to $w$.
The overall number of sets of candidate paths is fewer than $(3^{2g})^{4g^2} = 3^{8g^3}$. 
For a fixed choice of such a set of paths (one path for each admissible pair $(v,w) \in L\times R$),
we want to test, within some given box $D_{i,j}$, whether all the $< 4g^2$ chosen paths are the shortest paths between the corresponding pairs of positions.
As unfolded next, we will apply this test for all boxes $D_{i,j}$,
and output those boxes at which the outcome is positive (for the current chosen set of shortest paths).
We will repeat the procedure for all $< 3^{8g^3}$ possible sets of candidate paths $P_{v,w}$.
Since we enumerated the staircase paths in lexicographical order earlier, we can easily proceed through all sets of candidate paths, using this order.

\paragraph{Testing a Fixed Choice of Shortest Paths.}
For each subsequence $A_i$, we create a (blue) point $\alpha_i$, and for each subsequence $B_j$ we create a (red) point $\beta_j$,
such that, for every admissible pair $(v,w)\in L\times R$, we have one coordinate for each path $P'_{v,w}\in S(v,w)$, different from the chosen path $P_{v,w}$.
The value of $\alpha_i$ (resp., $\beta_j$) at that coordinate is the corresponding expression~(\ref{alpha}) (resp.,~(\ref{beta})).
The points $\alpha_i$ and $\beta_j$ are embedded in $\reals^{d_g}$, 
where $d_g = \sum_{(v,w)}{\Gamma_{v,w}}$
is the sum over all admissible pairs $(v,w)\in L\times R$, and $\Gamma_{v,w}$ is the number of monotone staircase paths from $v$ to $w$ minus $1$.
As discussed earlier, $d_g = O(3^{2g})$.

We have that a (blue) point
\[
\alpha_i = \left(\ldots,A_i(P^{\row}_{v,w}(1)) + \cdots + A_i(P^{\row}_{v,w}(r)) - A_i(P'^{\row}_{v,w}(1)) - \cdots - A_i(P'^{\row}_{v,w}(t)), \dots \right)
\]
is dominated by a (red) point
\[
\beta_j = \left(\ldots, B_j(P^{\col}_{v,w}(1)) + \cdots + B_j(P^{\col}_{v,w}(r)) - B_j(P'^{\col}_{v,w}(1)) - \cdots - B_j(P'^{\col}_{v,w}(t)), \ldots \right),
\]
if and only if each of the paths that we chose (a path for every admissible pair $(v,w)\in L\times R$) is the shortest path between the corresponding positions $v,\,w$ in box $D_{i,j}$.
The number of points is $2s = \Theta(n/g)$, and the time to prepare them, i.e., to compute all their coordinates,
is $O(2s \cdot 3^{2g} \cdot g) = O(3^{2g} n)$.

By Lemma~\ref{lem:redblue}, we can report all pairs of points $\left(\alpha_i , \beta_j \right)$ such that $\alpha_i$ is dominated by $\beta_j$, in 
$O\left( c^{O(3^{2g})}_\eps (n/g)^{1+\eps} + K\right)$ time, where $K$ is the number of boxes at which the test of our specific chosen paths comes out positive.
As mentioned earlier, we use $\eps =1/2$, with $c_{\eps} \approx 3.42$.

This runtime is for a specific choice of a set of shortest paths between all admissible pairs in $L \times R$.
As already mentioned,
we repeat this procedure at most $3^{8g^3}$ times.
Overall, we will report exactly $s^2 = \Theta\left((n/g)^2\right)$ dominating pairs (red on blue),
because the set of shortest paths between admissible pairs in $L\times R$ in each box $D_{i,j}$ is unique
(recall that we assumed that any pair of distinct staircase paths in a box do not have the same cost).
Since the overall number of sets of candidate paths is bounded by $3^{8g^3}$, one path for each admissible pair,
the overall runtime for all invocations of the bichromatic dominance reporting algorithm (including preparing the points) is
\[
O\left(3^{8g^3} \left( 3^{2g} n + c^{O(3^{2g})}_\eps (n/g)^{1+\eps} \right) + (n/g)^2 \right). \\
\]

Recall that, so far, we have assumed that all the differences within the absolute values
$D_{i,j}[\ell, m] = \bigl|A_i(\ell) - B_j(m) \bigr|$ are positive,
which allowed us to drop the absolute values, and write
$D_{i,j}[\ell, m]  = A_i(\ell) - B_j(m)$, for every $i, j \in [s]$, and $\ell, m \in [g]$,
thereby facilitating the use of the extended Fredman trick~\eqref{eq:Fredman}.
Of course, in general this will not be the case, so, in order to still be able to drop the absolute values, we also have to
verify the signs of all these differences.

For each box $D_{i,j}$, there is a {\em unique} sign assignment $\sigma^* : [g] \times [g] \rightarrow \{-1, 1\}$ such that
\[D_{i,j}[\ell, m] = \bigl| A_i(\ell) - B_j(m) \bigr| = \sigma^*(\ell, m)(A_i(\ell) - B_j(m)),\]
for every $\ell, m \in [g]$ (our ``general position'' assumption implies that each difference is nonzero).
Thus for any staircase path $P= \left(P^{\row}, P^{\col}\right)$ in $D_{i,j}$, of length $t^*$, we have
\[\cost_{i,j}(P) = \sum^{t^*}_{t=1} \sigma^*(P(t)) \left(A_i(P^{\row}(t)) - B_j(P^{\col}(t)) \right). \]

Now we proceed as before, testing sets of paths, but now we also test sign assignments of the box,
by trying {\em every} possible assignment $\sigma  :  [g] \times [g] \rightarrow \left\{-1,\, 1\right\}$,
and modify the points $\alpha_i$ and $\beta_j$, defined earlier, by
(i) adding sign factors to each term,
and (ii) adding coordinates that enable us to test whether $\sigma$ is the correct assignment $\sigma^*$ for the corresponding boxes $D_{i,j}$. 

Denote by $P$ a candidate for the shortest path for some admissible pair of positions $(v,w)\in L \times R$, and let $\sigma$ be a candidate sign assignment. 
Then, for every other path $P'\in S(v,w)$, 
we have the following modified coordinates for $\alpha_i$ and $\beta_j$ respectively.
\small 
\begin{align*}
&\left(\ldots,\sigma(P(1)) A_i(P^{\row}(1)) + \cdots + \sigma(P(r))A_i(P^{\row}(r)) - \sigma(P'(1))A_i(P'^{\row}(1))  - \cdots -  \sigma(P'(t)) A_i(P'^{\row}(t)), \dots \right), \\
&\left(\ldots,\sigma(P(1)) B_j(P^{\col}(1)) + \cdots + \sigma(P(r))B_j(P^{\col}(r)) - \sigma(P'(1)) B_j(P'^{\col}(1))  - \cdots -  \sigma(P'(t))B_j(P'^{\col}(t)), \ldots \right),
\end{align*}
\normalsize
where we use the same notations as in~(\ref{eq:abs}), (\ref{alpha}), and (\ref{beta}).
In addition, to validate the correctness of $\sigma$,
we extend $\alpha_i$ and $\beta_j$ by adding the following $g^2$ coordinates to each of them.
For every pair $(\ell,m) \in [g] \times [g]$, we add the following coordinates to $\alpha_i$ and $\beta_j$ respectively.
\begin{align*}
&\left(\ldots, -\sigma(\ell,m)A_i(\ell),\ldots \right), \\
&\left(\ldots,  -\sigma(\ell,m)B_j(m), \ldots \right).
\end{align*}
This ensures that a point $\alpha_i$ is dominated by a point $\beta_j$  iff
$D_{i,j}[\ell, m]  = \sigma(\ell, m) \left( A_i(\ell) - B_j(m) \right)$,
for every $\ell,m \in [g]$, and 
all the $< 4g^2$ candidate paths that we test are indeed shortest paths in $D_{i,j}$. 

The runtime analysis is similar to the preceding one, but now we increase the number of candidate choices by a factor of $2^{g^2}$ 
(this factor bounds the number of all possible sign assignments),
and the dimension of the space where the points are embedded increases by $g^2$ additional coordinates. 
We now have $2s = \Theta(n/g)$ points in $\reals^{d_g + g^2}$ ($d_g = O(3^{2g})$ is as defined earlier), and
the time to prepare them (computing the value of each coordinate) is $O((n/g) (d_g + g^2) g ) = O(3^{2g} n)$.
There are at most $3^{8g^3}$ sets of candidate paths to test, and for each set, there are at most $2^{g^2}$ sign assignment to test,
so in total, we invoke the bichromatic dominance reporting algorithm at most  $2^{g^2}3^{8g^3} < 3^{8g^3 + g^2}$ times,
for an overall runtime (including preparing the points) of
\[
O\left(3^{8g^3 + g^2}\left(3^{2g} n + c^{O(3^{2g}) + g^2}_\eps (n/g)^{1+\eps} \right) + (n/g)^2 \right).
\]

By setting $\eps= 1/2$ and $g = \delta \log\log n$, for a suitable sufficiently small constant $\delta$, the first two terms become negligible (strongly subquadratic), and the runtime is therefore dominated by the output size,
that is $O\left( (n/g)^2 \right) = O\left(n^2 / (\log\log n)^2 \right)$.
Each reported pair $\left(\alpha_i , \beta_j \right)$ certifies that the current set of $ < 4g^2$ chosen candidate paths are all shortest paths in box $D_{i,j}$.
Each of the $s^2 = \Theta\left((n/g)^2 \right)$ sets of shortest paths is represented by $O(g^3) = O((\log\log n)^3)$ bits
(there are $< 4g^2$ shortest paths connecting admissible pairs, each of length at most $2g-1$, and each path can be encoded by its first position, followed by the sequence of its at most $2g-2$ moves, where each move is in one of the three directions up/right/up-right),
and thus it can easily be stored in one machine word (for sufficiently small $\delta$).
Moreover, we have an order on the pairs $(v,w)$ (induced by our earlier enumeration), so for each set, we can store its shortest paths in this order,
and therefore, accessing a specific path (for some admissible pair) from the set takes $O(1)$ time (in the word-RAM model that we assume). 

Note, however, that we obtain only the {\em positions} that the paths traverse and not their {\em cost}.
In later stages of our algorithm we will also need to compute, on demand, the cost of certain paths,
but doing this naively would take $O(g)$ time per path, which is too expensive for us.
To handle this issue, when we choose a candidate sign assignment $\sigma$,
and a set $S$ of the $< 4g^2$ paths as candidates for the shortest paths,
we also compute and store, for each path $P\in S$ that we have not yet encountered, the {\em rows-cost} of $P$ in $A_i$,
\[ V^{\row}_i(P, \sigma) =  \sigma(P(1))A_i(P^{\row}(1)) + \cdots + \sigma(P(t^*))A_i(P^{\row}(t^*)),\]
for every $i\in [s]$, and the {\em columns-cost} of $P$ in $B_j$,
\[ V^{\col}_j(P, \sigma) = \sigma(P(1))B_j(P^{\col}(1)) + \cdots + \sigma(P(t^*))B_j(P^{\col}(t^*)),\]
for every $j\in [s]$, where $t^*$ is the length of $P$.  
Observe that, for the correct sign assignment $\sigma^*$ of box $D_{i,j}$,
\begin{equation}\label{eq:cost}
\cost_{i,j}(P) = V^{\row}_i(P, \sigma^*) - V^{\col}_j(P, \sigma^*).
\end{equation}
We do not compute  $V^{\row}_i(P, \sigma) - V^{\col}_j(P, \sigma)$ yet, but
only compute and store (if not already stored) the separate quantities $V^{\row}_i(P, \sigma)$ and $V^{\col}_j(P, \sigma)$, for each $P\in S$, for every chosen set $S$, and sign assignment $\sigma$.
We store the values $V^{\row}_i(P, \sigma)$ and $V^{\col}_j(P, \sigma)$ in arrays, ordered by the earlier enumeration of all staircase paths,
so that given a staircase path $P$, and indices $i, j \in \left[\frac{n}{g-1}\right]$, we can retrieve, upon demand, the values $V^{\row}_i(P, \sigma^*)$ and $V^{\col}_j(P, \sigma^*)$,
and compute $\cost_{i,j}(P)$ by using~(\ref{eq:cost}), in $O(1)$ time.
In total, over all possible candidate paths and sign assignments, this takes $O(2^{g^2}3^{2g} \cdot (n/g)\cdot g) = O(3^{g^2 + 2g} n)$ time and space, 
which is already subsumed by the time (and space) bound for reporting dominances from the previous stage.

To summarize this stage of the algorithm, we presented a subquadratic-time preprocessing procedure, which runs in $O\left( (n/g)^2 \right) = O\left(n^2 / (\log\log n)^2 \right)$ time,
such that for any box $D_{i,j}$, and an admissible pair of positions $(v,w)\in L\times R$,
we can retrieve the shortest path $P^*_{v,w}$ in $O(1)$ time,
and can also compute $\cost_{i,j}(P^*_{v,w})$ in $O(1)$ time. This will be useful in the next stage of our algorithm.

\subsection*{Second Stage: Compact Dynamic Programming}
\begin{figure}[t]
	\centering
	\includegraphics[scale = 0.6]{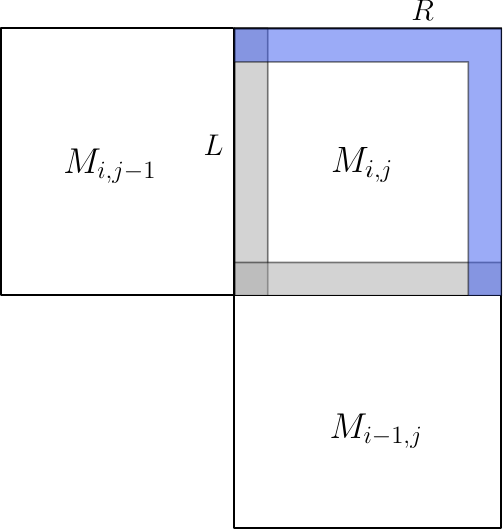}
	\caption{The $L$-boundary (shaded in gray) of box $M_{i,j}$ overlaps with the top boundary of $M_{i-1,j}$ and the right boundary of $M_{i,j-1}$. 
		Once we have the values of $M$ at the positions of the $L$-boundary of $M_{i,j}$, our algorithm computes the values of $M$ at the positions of its $R$-boundary (shaded in blue). }
	\label{fig:matrix}
\end{figure}

Our approach is to view the $(n+1) \times (n+1)$ matrix $M$ from the dynamic programming algorithm (see Section~\ref{sec:quadratic})
as decomposed into $s^2 = \left( \frac{n}{g-1} \right)^2$ boxes $M_{i,j}$, each of size $g\times g$,
so that each box $M_{i,j}$ occupies the same positions as does the corresponding box $D_{i,j}$.
That is, the indices of the rows (resp., columns) of $M_{i,j}$ are those of $A_i$ (resp., $B_j$). In particular,
for each $i,j\in [s]$,
the positions $\left(\cdot, g \right)$ on the right boundary of each box $M_{i,j}$ coincide
with the corresponding positions $\left(\cdot, 1 \right)$ on the left boundary of $M_{i,j+1}$,
and the positions $\left(g, \cdot \right)$ on the top boundary of $M_{i,j}$ coincide with the corresponding positions $\left(1, \cdot \right)$ on the bottom boundary of $M_{i+1, j}$. 
Formally, 
$M_{i,j}[\ell, m]= M\left[(i-1)(g-1) + \ell,\, (j-1)(g-1) + m \right]$,
for each position $(\ell, m) \in [g] \times [g]$. See Figure~\ref{fig:matrix} for an illustration.

Our strategy is to traverse the boxes, starting from the leftmost-bottom one $M_{1,1}$, where
we already have the values of $M$ at the sequence $L$ of positions of its left and bottom boundaries (initialized to the same values as in the algorithm in Section~\ref{alg:DTWquadratic}), and we compute the values of $M$ on its top and right boundaries $R$.
We then continue to the box on the right, $M_{1,2}$, now having the values on its $L$-boundary
(where its left portion overlaps with the $R$-boundary of $M_{1,1}$ and its bottom portion is taken from the already preset bottom boundary),
and we compute the values of $M$ on its $R$-boundary.
We continue in this way until we reach the rightmost-bottom box $M_{1, s}$.
We then continue in the same manner in the next row of boxes, starting at $M_{2,1}$ and ending at $M_{2,s}$, and keep going through the rows of boxes in order.
The process ends once we compute the values of $M$ on the $R$-boundary of the rightmost-top box $M_{s, s}$,
from which we obtain the desired entry $M[n, n]$.

For convenience, we enumerate the positions in $L$ as $L(1), \ldots, L(2g-1)$ in ``clockwise'' order, so that $L(1)$ is the rightmost-bottom position $(1,g)$,
and $L(2g-1)$ is the leftmost-top position $(g,1)$. Similarly, we enumerate the positions of $R$ by $R(1), \ldots, R(2g-1)$ in ``counterclockwise'' order,
with the same starting and ending locations.
Let $M_{i,j}(L) = \{ M_{i,j}[L(1)], \ldots M_{i,j}[L(2g-1)] \}$ and  $M_{i,j}(R) = \{  M_{i,j}[R(1)], \ldots M_{i,j}[R(2g-1)] \}$, for $i,j \in [s]$.

By definition, for each position $(\ell, m)\in [n+1]\times [n+1]$, $M[\ell, m]$ is the minimal cost of a staircase path from $(0,0)$ to $(\ell, m)$.
It easily follows, by construction, that for each box $D_{i,j}$, and for each position $w \in R$, we have
\begin{equation}\label{eq:cumulative-cost}
M_{i,j}[w] = \min_{\substack{v \in L\\ (v,w) \text{ admissible}}} \Bigl\{ M_{i,j}[v] + \cost_{i,j}(P^*_{v,w})  \Bigr\}.
\end{equation}
(Note that, by definition, the term $D_{i,j}[v]$ is included in $M_{i,j}[v]$ and not in $\cost_{i,j}(P^*_{v,w})$, so it is not doubly counted.)
For each box $M_{i,j}$ and each position $w \in R$, our goal is thus to compute
the position $u \in L$ that attains the minimum in~(\ref{eq:cumulative-cost}), and the corresponding cost $M_{i,j}[w]$.
We call such $(u,w)$ the {\em minimal pair} for $w$ in $M_{i,j}$.

For each box $D_{i,j}$, and each admissible pair $(v,w)\in L \times R$,
we refer to the value $M_{i,j}[v] + \cost_{i,j}(P^*_{v,w})$ as the {\em cumulative cost} of the pair $(v,w)$, and denote it by $\ccost(v,w)$.

We can rewrite~(\ref{eq:cumulative-cost}), for each position $w \in R$, as
\[
M_{i,j}[w] = \min \bigl\{ M^W_{i,j}[w],\, M^S_{i,j}[w] \bigr\},
\]
where $M^S_{i,j}[w]$ is the minimum in~(\ref{eq:cumulative-cost}) computed only over $v \in \left\{ L(1), \ldots, L(g)  \right\}$,
which is the portion of $L$ that overlaps the $R$-boundary of the bottom (south) neighbor $M_{i-1, j}$ (when $i > 1$),
and $M^W_{i,j}[w]$ is computed over $v \in \left\{ L(g), \ldots, L(2g-1)  \right\}$, which overlaps the $R$-boundary of the left (west) neighbor $M_{i,j-1}$ (when $j >1$).
See Figure~\ref{fig:matrix} for a schematic illustration.
(Recall that the bottommost row and the leftmost column of $M$ are initialized with $\infty$ values, except their shared cell $M[0,0]$ that is initialized with $0$.)
The output of the algorithm is $M_{s,s}[R(g)] = M_{s,s}[g,g] = M[n, n]$.
We can also return the optimal coupling, by using a simple backward pointer tracing procedure, similar in principle to the one mentioned for the quadratic algorithm in Section~\ref{sec:quadratic}.

\paragraph{Computing Minimal Pairs.}
We still have to explain how to compute the minimal pairs $(u,w)$ in each box $M_{i,j}$.
Our preprocessing stage produces, for every box $D_{i,j}$, the set of all its shortest paths $S_{i,j}=\{ P^*_{v,w} \mid  (v,w) \in L \times R \}$
(ordered by the earlier enumeration of $L \times R$ and including only admissible pairs), and we can also retrieve the cost of each of these paths in $O(1)$ time (as explained earlier in the preprocessing stage).
The cumulative cost (defined above) of each such pair $(v,w)$ can also be computed in $O(1)$ time, assuming we have already computed $M_{i,j}[v]$.
A naive, brute-force technique for computing the minimal pairs is to compute all the cumulative costs $\ccost_{i,j}(v,w)$,
for all admissible pairs $(v,w) \in L\times R$, and select from them the minimal pairs.
This however would take $O(g^2)$ time for each of the $s^2$ boxes, for a total of $\Theta(g^2 s^2) = \Theta(n^2)$ time, which is what we want to avoid.

Fortunately, we have the following important lemma, which lets us compute all the minimal pairs within a box,
significantly faster than in $O(g^2)$ time.

\begin{figure}[t]
	\centering
	\includegraphics[scale = 0.5]{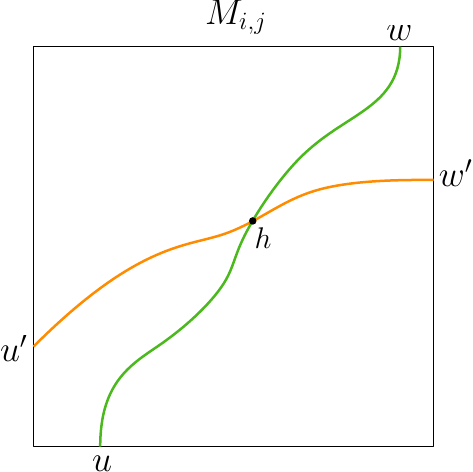}
	\caption{By Lemma~\ref{lem:monge}, if $(u,w)$ and $(u',w')$ are minimal pairs in $M_{i,j}$, then the illustrated scenario is impossible,
		since the path $P^*_{u,w}$ (in green) is a portion of the shortest path from $M[0,0]$ to $M_{i,j}[w]$,
		and the path $P^*_{u',w'}$ (in orange) is a portion of the shortest path from $M[0,0]$ to $M_{i,j}[w']$.
		The illustrated intersection implies that one of the latter paths can decrease its cumulative cost by replacing its portion that ends at $h$
		by the respective portion that ends in $h$ of the other path (recall that we assume that there are no two paths with the same cost), which contradicts the fact that both of these paths are shortest paths.}
	\label{fig:monge}
\end{figure}

\begin{lemma}\label{lem:monge}
	For a fixed box $D_{i,j}$, and for any pair of distinct positions $w, w' \in R$,
	let $u, u' \in L$ be the positions for which $(u,w)$ and $(u',w')$ are minimal pairs in $M_{i,j}$.
	Then their corresponding shortest paths $P^*_{u,w}$ and $P^*_{u',w'}$ can partially overlap but can never cross each other.
	Formally, assuming that $w > w'$ (in the counterclockwise order along $R$), we have that for any $\ell, \ell', m \in [g]$, if $(\ell, m) \in P^*_{u,w}$ and $(\ell', m) \in P^*_{u',w'}$ then $\ell \geq \ell'$.
	That is, $P^*_{u,w}$ lies fully above $P^*_{u',w'}$ (partial overlapping is possible).
	In particular, we also have $u \geq u'$ (in the clockwise order along $L$).
\end{lemma}
Lemma~\ref{lem:monge} asserts the so-called {\em Monge property} of shortest-path matrices (see, e.g.,~\cite{B96, KMNS12}).
See Figure~\ref{fig:monge} for an illustration (of an impossible crossing) and a sketch of a proof.

Using Lemma~\ref{lem:monge}, we first present a divide-and-conquer paradigm for computing 
the minimal pairs within a box $M_{i,j}$ in $O(g\log g)$ time, which is conceptually simple to perceive. 
However, this is not the best we can do. Afterwards, we present an even more efficient procedure that takes only $O(g)$ time in total. 

We start by setting the median index $k = \floor{|R| /2}$ of $|R|$,
and compute the minimal pair $(u, R(k))$ and its $\ccost(u, R(k))$, naively, in $O(g)$ time, as explained above.
The path $P^*_{u, R(k)}$ decomposes
the box $M_{i,j}$ into two parts, so that
one part, $X$, consists of all the positions in $M_{i,j}$ that are (weakly) {\em above} $P^*_{u,R(k)}$,
and the other part, $Y$, consists of all the positions in $M_{i,j}$ that are (weakly) {\em below} $P^*_{u,R(k)}$,
so that $X$ and $Y$ are disjoint, except for the positions along the path $P^*_{u,R(k)}$ which they share.
By Lemma~\ref{lem:monge}, the shortest paths between any other minimal pair of positions in $L \times R$ can never cross $P^*_{u, R(k)}$.
Thus, we can repeat this process separately in $X$ and in $Y$.
Note that the input to each recursive step is just the sequences of positions of $X$ and $Y$ along $L$ and $R$, respectively (and we encode each sequence simply by its first and last elements); there is no need to keep track of the corresponding portion of $M_{i,j}$ itself.

Denote by $T(a, b)$ the maximum runtime for computing all the minimal pairs $(u,w)$, within any box $M_{i,j}$,
for $u$ in some contiguous portion $L'$ of $a$ entries of $L$, and $w$ in some contiguous portion $R'$ of $b$ entries of $R$.
Clearly,  $T(1, b) = O(b)$, and $T(a, 1) = O(a)$. In general, the runtime is bounded by the recurrence
\[
T(a, b) = \max_{k \in [a]}\,\Bigl\{\, T(k,\, \floor{b/2})  + T(a-k+1,\, \floor{b/2})\,\Bigr\} +O(a).
\]
It is an easy exercise to show, by induction, that the solution of this recurrence satisfies $T(a,b) = O\left((a+b)\log b \right)$.
Thus, the runtime of the divide-and-conquer procedure described above, for a fixed box $M_{i,j}$, is $O\left((|R| + |L|)\log|R| \right) = O(g\log g)$. 

The runtime of computing $M_{i,j}(R)$ for all $s^2 = \Theta\left((n/g)^2 \right)$ boxes is thus
$O\left( (n/g)^2 g\log g \right) = O\left( n^2\log g /g \right)$. 
Overall, including the preprocessing stage, the total runtime of the algorithm is
$O\left( (n/g)^2 + n^2\log g /g\right) = O\left(n^2\log g /g \right)$.
As dictated by the preprocessing stage, we need to choose $g = \Theta(\log\log n)$, so the overall runtime is
$O\left( n^2 \log\log\log n / \log\log n \right)$.

\paragraph{A Further Improvement: Removing the $\log\log\log n$ Factor.}
We can speed up the computation of minimal pairs even further, so that computing $M_{i,j}(R)$ for each box will take $O(g)$ time, improving the $O(g\log g)$ bound of the divide-and-conquer algorithm described above.

For each box $M_{i,j}$, denote by $M^{LR}_{i,j}$ the $(2g-1)\times (2g-1)$ matrix such that the $L$ and $R$ positions of $M_{i,j}$ correspond to the rows and columns of $M^{LR}_{i,j}$, respectively. Namely, each pair $(v,w)\in L\times R$ corresponds to a cell in $M^{LR}_{i,j}$ that its value is $\ccost_{i,j}(v,w)$ (that is, the cost of the shortest path from the origin of $M$ that goes through $v$ and ends at $w$).
For convenience, we denote by $M^{LR}_{i,j}[\ell, m]$ the cell that corresponds to the pair $(L(\ell), R(m)) \in L\times R$. 

Lemma~\ref{lem:monge} implies the following observation.

\begin{observation}\label{obs:monge-matrix}
	The matrix $M^{LR}_{i,j}$ is a Monge matrix. 
	That is, for every $\ell < \ell' \in [2g-1]$ and every $m < m' \in [2g-1]$, we have that
	\begin{equation}\label{eq:monge-matrix}
	M^{LR}_{i,j}[\ell, m] + M^{LR}_{i,j}[\ell', m'] < M^{LR}_{i,j}[\ell, m'] + M^{LR}_{i,j}[\ell', m].
	\end{equation}
\end{observation}
Indeed, it is easy to check that if Equation~(\ref{eq:monge-matrix}) does not hold then we have a contradiction to Lemma~\ref{lem:monge}.
(See~\cite{BKR96} for a survey on Monge matrices and their applications.)

Observation~\ref{obs:monge-matrix} immediately implies that the matrix $M^{LR}_{i,j}$ is {\em totally monotone}. 
That is, for every $\ell < \ell' \in [2g-1]$ and every $m < m' \in [2g-1]$, we have that 
\begin{equation*}
M^{LR}_{i,j}[\ell, m] > M^{LR}_{i,j}[\ell, m'] \implies M^{LR}_{i,j}[\ell', m] > M^{LR}_{i,j}[\ell', m'].
\end{equation*}

To compute $M_{i,j}(R)$ we need to find the minimum of every column $w\in R$ (i.e., to find $\min_{v\in L}{M^{LR}_{i,j}[v,w]}$).
Since $M^{LR}_{i,j}$ is totally monotone, we can use the SMAWK algorithm~\cite{SMAWK} to compute the minimum of each column of $R$ in total $O(|L|+|R|) = O(g)$ time.

Thus, the runtime of computing $M_{i,j}(R)$ for all $s^2 = \Theta\left((n/g)^2 \right)$ boxes becomes $O\left( n^2 /g \right)$. 
This bound in fact dominates the total runtime of the algorithm, provided the we choose $g = \Theta(\log\log n)$, due to the preprocessing stage.  
Hence, we obtain that the total runtime of the algorithm is 
$O\left(n^2 / \log\log n \right)$.

This completes the proof of Theorem~\ref{thm:first} for \DTW{} on a pair of point-sequences in $\reals$. \qed

\subsection{Extension to High-Dimensional Polyhedral Metric Spaces}\label{sec:extended}
The algorithm described above can be extended to work in higher dimensional spaces $\reals^d$, for any constant $d$, when the underlying metric is polyhedral.
That is, the underlying metric is induced by a norm, whose unit ball is a symmetric convex polytope with $O(1)$ facets.
To illustrate this extension,
consider the $L_1$-metric in $\reals^d$, whose unit ball is the symmetric cross-polytope $|x_1|+ \cdots + |x_d| \leq 1$, with $2^d$ facets.
In this case, each entry in the blocks $D_{i,j}$ is a sum of $d$ absolute values.
By choosing a candidate sign assignment for all the absolute values, 
each comparison that the algorithm faces is a sign test of a $2d$-linear expression in the input (with coefficients $1, -1$),
and the extended~Fredman~trick~\eqref{eq:Fredman} can then be applied when comparing the costs of two staircase paths.
Then, in much the same way as before, we can encode the inequalities into red and blue points $\alpha_i$ and $\beta_j$,
and use a suitable modification of the preceding machinery
to compare costs of staircase paths and validate sign assignments correctness.
Omitting further details, we get a subquadratic algorithm for \DTW{} in such a higher-dimensional setup under the $L_1$-metric,
with the same asymptotic time bound as that of the algorithm described above,
but with the constant of proportionality depending (exponentially) on $d$.

To handle general polyhedral metrics, let $K$ denote the unit ball of the metric.
For each pair of points $p_\ell \in A$, $q_m \in B$,
we choose some facet of $K$ as a candidate for the facet that is hit by the oriented ray that emanates from the origin in the direction of the vector $\overrightarrow{p_\ell q_m}$
(this replaces the sign assignments used in the one-dimensional case and for the $L_1$-metric).
Given such a candidate facet, $\dist(p_\ell,q_m)$ is a linear expression, and the extended Fredman trick, with all the follow-up for comparing costs of staircase paths can be applied, except that we also need to validate the correctness of our chosen candidate facet of $K$.
This is done as follows.

Assume, without loss of generality, that each facet of $K$ is a $(d-1)$-simplex (this can be achieved by a suitable triangulation of the facets).
Consider a simplex-facet $f$, and let $F$ be the cone spanned by $f$ with apex at the origin.
$F$ is the intersection of $d$ halfspaces, each of the form $\left< h_i,x \right> \geq 0$, for suitable normal unit vectors $h_1,\ldots, h_d$.
In order to verify that the direction $\overrightarrow{p_\ell q_m}$ hits $f$, 
we need to verify that $\left< h_i, q_m - p_\ell \right> \geq 0$, or that $\left< h_i, q_m \right>\geq \left< h_i, p_\ell \right>$, for $i=1,\ldots, d$.
These are $d$ linear tests, which fit well into the frame of the extended Fredman trick (they replace the sign test that are used in the one-dimensional case, and in the $L_1$-case). 

Again, omitting the further, rather routine details, we obtain a subquadratic algorithm for \DTW{} in any fixed dimension, under any polyhedral metric,
with the same runtime as in Theorem~\ref{thm:first} and as stated in Theorem~\ref{thm:second}.
The constant of proportionality depends on the dimension $d$, and on the complexity of the unit ball $K$ of the metric (i.e., the number and complexity of its facets).

\subsection{Lifting the General Position Assumption}\label{subsec:lifting}
In the algorithm above, we assumed that in each box $D_{i,j}$ there are no two staircase paths with the same cost. 
This assumption was crucial for preserving the overall output size of the dominance reporting routines to be $O(n^2/g^2)$.
Specifically, all we need to ensure is that for each admissible boundary pair from $L\times R$,
there will be only one staircase path with minimum cost. Our goal is to be able to break ties consistently.
However, this is not trivial, as we must find a way to do it while using the Fredman-Chan mechanism. We can do it as follows.

In the preprocessing stage, our algorithm enumerated all the $< 3^{2g}$ staircase paths in a $g\times g$ grid.
These enumerations are done in a natural lexicographic order and thus induce a total order on the staircase paths.
Denote this total order by $\mathcal{L}$. (Note that $\mathcal{L}$ is independent of the values of $A$ and $B$.)

Let $A, B$ be two given input sequences of points (numbers) in $\reals$
(a similar solution works for the extension to $\reals^d$ under polyhedral metrics described above). 
First, sort $A$ and $B$ in increasing order in $O(n\log n)$ time.
Find a {\em positive closest pair} $(a,b)\in A\times B$, i.e., a pair satisfying
\[
|a-b| = \min_{(a_i, b_j)\in A\times B:\, |a_i-b_j| >0}\left\{|a_i - b_j|\right\}.
\]
This can be done while merging the sorted $A$ and $B$, in $O(n)$ time.
(If we are in a polyhedral $\reals^d$ metric space we use a straightforward modification of the standard $O(2^d n\log n)$ divide-and-conquer closest pair algorithm of Bentley and Shamos~\cite{Bentley76, Bentley80} to find the positive closest pair in the set $A\cup B$.)
Put $\eps = |a-b|$.  
For every boundary pair in $L\times R$ there are strictly fewer than $3^{2g}$ staircase paths, denote this number by $r$.
Set 
\[
\eps_1 = \eps/3^{2g} < \eps_2 = 2\eps/ 3^{2g} < \eps_3 =  3\eps/ 3^{2g} < \cdots < \eps_r = r\eps/ 3^{2g}.
\]

For every boundary pair $(v,w)\in L\times R$, and every staircase path $P_{v,w}$ (recall that $P_{v,w}$ is a sequence of positions in the $g\times g$ grid, and is independent of the values of $A$ and $B$),
we check for the index $k$ of $P_{v,w}$ in the total order $\mathcal{L}$, and add $\eps_k$ to $\cost_{i,j}(P_{v,w})$, for every $i,j\in \left[\frac{n}{g-1} \right]$.

For a boundary pair $(v,w)\in L\times R$, let $P_{v,w}$ and $P'_{v,w}$ be two distinct staircase paths,
and let $k, k'\in [r]$ be their corresponding (distinct) indices in the total order $\mathcal{L}$.
Assume, without loss of generality, that $k < k'$ (it must be that either $k < k'$ or $k' < k$, since the two paths are distinct, and $\mathcal{L}$ is a total order).
Since $\eps_k < \eps_{k'} < |a-b|$, it holds that for every $i,j\in \left[\frac{n}{g-1} \right]$,
\[
\cost_{i,j}(P_{v,w}) \leq \cost_{i,j}(P'_{v,w}) \text{\: if and only if\:} \cost_{i,j}(P_{v,w}) + \eps_k < \cost_{i,j}(P'_{v,w}) + \eps_{k'}.
\]

We now proceed with the same steps of the algorithm we described in Section~\ref{sec:dtw} (and~\ref{sec:extended}) but with the modified path costs.
(Note that we used the same $\eps_1, \ldots, \eps_r$ for all boxes $D_{i,j}$, thus we can still use the extended Fredman trick for the new costs.) 
By the above, ties on the original costs of (any) two distinct staircase paths break on their new costs, according to their order in $\mathcal{L}$, while the other relations ($<, >$) are preserved.

\section{Geometric Edit Distance in Subquadratic Time}\label{sec:ed}
In this section, we show how our \DTW{} algorithm from Section~\ref{sec:dtw} can be modified to compute $\mathsf{ged}(A,B)$ (and optimal matching).
Recall the definitions of monotone matching (see Figure~\ref{fig:matching}), $\mathsf{ged}(A,B)$, and optimal matching from Section~\ref{DTW:ProblemStatement}. 
First, we overview the standard dynamic programming algorithm for computing \GED{} between two sequences $A=(p_1,\ldots, p_n)$ and $B=(q_1, \ldots, q_{n})$,
each of $n$ points in $\reals$.

\paragraph{The Quadratic Time \GED{} Algorithm.}
\begin{description}\label{alg:quadratic-ged}
	\addtolength{\itemsep}{-0.5\baselineskip}
	\item[1.$\;$] Initialize an $(n+1)\times (n+1)$ matrix $M$ and set $M[0,0]:=0$.
	\item[2.$\;$] For each $\ell \in [n]$
	\item [2.1.$\;\;$] $M[\ell, 0]:=  \ell\rho$, $M[0, \ell]:= \ell\rho$. 
	\item[3.$\;$] For each $\ell\in [n]$, 
	\item[3.1.$\;\;$] For each $m\in [n]$,
	\item[3.1.1$\;\;\;$] 	$M[\ell,m] :=  \min \Bigl\{M[\ell-1,m] + \rho,\, M[\ell,m-1] + \rho,\, M[\ell-1, m-1] + \bigl|p_\ell - q_m\bigr| \Bigr\}$.
	\item[4.$\;$] Return $M[n,n]$.
\end{description}
The optimal matching can be retrieved by maintaining pointers from each $(\ell, m)$ to the preceding position
$(\ell', m')\in \left\{ (\ell-1, m),\, (\ell, m-1),\, (\ell-1, m-1) \right\}$ through which $M[\ell , m]$ is minimized.
By tracing these pointers backwards from $(n, n)$ to $(0, 0)$ and including in the matching only the positions that we reach ``diagonally'' (when going backwards), we obtain the optimal matching.

\paragraph{Subquadratic Time \GED{} Algorithm.}
Recall the all-pairs-distances matrix $D$ and its decomposition into boxes $D_{i,j}$, as defined in Section~\ref{sec:dtw}.
For a monotone matching $\mathcal{M}$ between two point-subsequences $A_i, B_j$, let $\cost_{i,j}(\mathcal{M})$ be the corresponding sum of distances in the definition of $\mathsf{ged}(A_i,B_j)$.
To adapt our \DTW{} algorithm for \GED{}, we modify the way we evaluate the cost of a staircase path $P$ in a box $D_{i,j}$,
so that it equals the cost of its corresponding monotone matching $\mathcal{M}(P)$ (defined below).

We view each box $D_{i,j}$ as a weighted directed grid graph $G$, whose vertices are the pairs of $[g]\times [g]$,
and its set of edges is
\begin{align*}
&\bigl\{\left<(\ell, m), (\ell + 1, m)\right> \mid \ell \in [g-1],\; m\in [g]  \bigr\} \\
\mathsmaller{\bigcup} &\bigl\{\left<(\ell, m), (\ell, m+1)\right> \mid \ell \in [g],\; m\in [g-1]  \bigr\} \\
\mathsmaller{\bigcup} &\bigl\{\left<(\ell,m), (\ell+1, m+1)\right> \mid \ell, m \in [g-1] \bigr\}.
\end{align*}
We refer to the edges in the first subset as vertical edges, the edges in the second subset as horizontal edges, and the ones in the third subset as diagonal edges.
The weight of the vertical and horizontal edges is set to $\rho$, and the weight of each diagonal edge $\left<(\ell,m), (\ell+1, m+1)\right>$ is $|A_i(\ell) - B_j(m)|$.
Each staircase path $P$ in $D_{i,j}$ is then a path in the graph $G$, whose corresponding monotone matching $\mathcal{M}(P)$ is defined to
consist of exactly all the pairs of points $(A_i(\ell), B_j(m))$ that correspond to the positions $(\ell, m)$
from the diagonal edges $\left<(\ell,m), (\ell+1, m+1)\right>$ of the path.

By defining $\cost_{i,j}(P)$ to be the weight of its corresponding path in $G$,
we obtain that $\cost_{i,j}(P) = \cost_{i,j}(\mathcal{M})$,
and that the dynamic programming matrix $M$ (given above) satisfies that
for each position $(\ell, m)\in [n+1]\times [n+1]$, $M[\ell, m]$ is the minimal cost of a staircase path from $(0,0)$ to $(\ell, m)$ in $D$.
This implies that Lemma~\ref{lem:monge} can be used in this setup too, for computing the values on the $R$-boundaries of the boxes $M_{i,j}$, as done in the second stage of our $\DTW{}$ algorithm.
Thus, once we have a corresponding data structure from the preprocessing procedure, we can apply the second stage of our $\DTW{}$ algorithm verbatim.

As for the preprocessing procedure,
the cost of a staircase path in a box $D_{i,j}$ is now a sum of distances $|A_i(\ell) - B_j(m)|,\, \ell, m\in [g]$, plus a multiple of the parameter $\rho$. 
Since $\rho$ is a fixed real number and the multiple of $\rho$ in the cost of a staircase path in $D_{i,j}$ only depends on the positions of the path (and is independent of the actual values of $A$ and $B$), we can execute a similar machinery as described in Section~\ref{sec:dtw}. That is, we can choose a candidate sign assignment as before, 
get a linear expression in $A_i(\ell)$ and $B_j(m)$ (which also involves a fixed multiple of $\rho$), then, the extended Fredman trick~\eqref{eq:Fredman} can be applied when comparing the costs of two staircase paths and validating the correctness of candidate sign assignments.
(Our algorithm works also for more general gap penalty functions, as long as they are linear in the coordinates of the points of $A\cup B$.)
The rest of the preprocessing procedure and the extension to high-dimensional polyhedral metric spaces are
similar to those we showed for \DTW{}.
In order to lift the general position assumption, a tiny modification to what is described in Section~\ref{subsec:lifting} is required; to set $\eps$ from Section~\ref{subsec:lifting} as the
minimum over the distance of the positive closest pair from $A\cup B$ and $\rho$, the rest is verbatim.

From the above, we obtain that $\mathsf{ged}(A,B)$ (and an optimal matching) can be computed in $O(n^2 / \log\log n)$ time,
as stated in Theorems~\ref{thm:first} and~\ref{thm:second} for \GED{}. \qed

\section*{Acknowledgments}
The first author would like to thank one of the anonymous reviewers of his PhD thesis,
for suggesting that we can also shave-off the $\log\log\log n$ factor from our algorithms that appear in previous versions of this paper (see~\cite{GS17-DTW}), using the SMAWK algorithm~\cite{SMAWK} for totally monotone matrices.

\bibliographystyle{abbrv}
\bibliography{DTW}

\end{document}